\documentclass[12pt]{iopart}
\usepackage{iopams}  

\usepackage{bm}
\usepackage{graphicx}
\usepackage{cite}

\newcommand{\mnp}{\mathnormal{\Psi}}

\newcommand{\sech}{\,\mathrm{sech}}

\begin{document}

\title{Splitting and focusing of neutrino collective states}   

\author{Mattias Marklund\dag\footnote[2]{E-mail:
  marklund@elmagn.chalmers.se},  
  Padma K.\ Shukla\S{$||$} and Lennart Stenflo$||$}

\address{\dag\ Department of Electromagnetics, Chalmers University of
Technology, SE--412 96 G\"oteborg, Sweden}

\address{\S\ Institut f\"ur Theoretische Physik IV, Fakult\"at f\"ur
  Physik und Astronomie, Ruhr-Universit\"at Bochum, D--44780 Bochum,
  Germany}

\address{$||$ Department of Plasma Physics, Ume{\aa} University,
  SE--901 87 Ume{\aa}, Sweden}

\date{\today}

\begin{abstract}
It is shown that the collective nonlinear interactions between intense 
neutrino or anti-neutrino fluxes and a dense neutrino plasma are
governed by a multi-dimensional 
coupled cubic Schr\"odinger equation in which the interaction
potential is positive or negative depending on the neutrino type. The
cubic Schr\"odinger equation 
describes the splitting and focusing of intense neutrino beams due to 
the nonlinear excitations associated with  
the modifications of the individual neutrino energies in a dense
neutrino background.   
\end{abstract}


\section*{}
Neutrinos and anti-neutrinos are produced in the Sun and in supernovae 
during violent processes. The propagation of intense neutrinos in 
ionised media (dense plasmas) is a topic of current interest in view
of the fact that neutrinos may be massive \cite{Ahmad}. Because of
their weak interaction with charged and neutral particles, neutrinos
(or anti-neutrinos) can travel great distances without being affected 
appreciably by material obstacles. Massive neutrinos can account for 
neutrino oscillations and associated flavour change in solar 
plasmas \cite{John,Raffelt}, while in supernovae and cosmology they 
may play a decisive role in reviving shocks for supernovae explosions,  
as well as in accounting for the  missing dark matter in our Universe
\cite{Dolgov}. 

In the present paper, we employ a semi-classical approach to 
consider the nonlinear interaction between neutrinos  
and anti-neutrinos in neutrino plasmas.  
Using an effective field theory approach, a system of coupled
nonlinear   
Schr\"odinger equations is derived, and it is shown that these
equations  
admit splitting and focusing of intense neutrino (or anti-neutrino)
beams due to the nonlinearities associated with the weak nuclear
forces of the  
neutrinos/anti-neutrinos that are interacting with the background
neutrinos.  

Suppose a single neutrino (or anti-neutrino) moves in a
neutrino--anti-neutrino  
admixture. The energy $E$ of the neutrino (anti-neutrino) is then
given by (see,   
e.g. \cite{Bethe,Kuo-Pantaleone,Silva-etal})
\begin{equation}\label{energy}
E = \sqrt{p^2c^2 + m^2c^4} + V_{\pm}(\bm{r},t) ,
\end{equation}
where $\bm{p}$ is the neutrino (anti-neutrino) momentum, $c$ is the
speed of light in vacuum, and $m$ is the neutrino mass. The effective
potential for a neutrino moving on a background of it's own flavor is
\cite{Kuo-Pantaleone,Notzhold-Raffelt} (see also
\cite{Weldon,Nunokawa-etal})  
$
V_{\pm}(\bm{r},t) = \pm 2\sqrt{2}G_F (n - \bar{n}),
$ 
while the potential for a neutrino moving on a background of a
different flavour takes the form 
$
V_{\pm}(\bm{r},t) = \pm \sqrt{2}G_F (n - \bar{n}).
$ 
Here $G_F/(\hbar c)^3  \approx 1.2\times 10^{-5}\,
\mathrm{GeV}^{-2}$, where $G_F$ is the Fermi constant, $n$ ($\bar{n}$) 
is the density of the background neutrinos (anti-neutrinos), and the $+$
($-$) is for the propagating neutrino (anti-neutrino). From
Eq.~(\ref{energy}) we can, using standard techniques \cite{Hasegawa},
obtain a Schr\"odinger equation for a neutrino (anti-neutrino) wave packet
${\mnp}(\bm{r}, t)$ \cite{Marklund-etal} 
(see also Ref.\ \cite{Tsintsadze-etal}  
for a similar treatment of neutrino--electron interactions)
\begin{equation}\label{nlse}
  i\frac{\partial{\mnp}}{\partial t} +
  \frac{\hbar}{2m_{\mathrm{eff}}}\nabla^2{\mnp} -
  \frac{V_{\pm}}{\hbar}{\mnp} = 0 ,
\end{equation}
where $m_{\mathrm{eff}} = m \gamma$, $\gamma = (1 - v^2/c^2)^{-1/2}$, and
$v$ is the magnitude of the group velocity of the neutrino wave packet. 

In the case of self-interacting neutrinos and anti-neutrinos, moving
with the same group velocity, we have $ n =
\langle|{\mnp}_{+}|^2\rangle$ and $\bar{n} =
\langle|{\mnp}_{-}|^2\rangle$,   
where ${\mnp}_{+}$ and ${\mnp}_{-}$ are the neutrino and
anti-neutrino wave functions,
respectively, and the angular bracket denotes the ensemble average. 
Thus, in this case Eq.\ (\ref{nlse}) turns into the coupled system   
\begin{eqnarray}
  i\frac{\partial{\mnp}_{\pm}}{\partial t} +
  \frac{\alpha}{2}\nabla^2{\mnp}_{\pm} \mp
  \beta(\langle|{\mnp}_{+}|^2\rangle - \langle|{\mnp}_{-}|^2\rangle
  ){\mnp}_{\pm} = 0 ,   
\label{interaction}
\end{eqnarray}
where $\alpha = \hbar/m_{\mathrm{eff}}$ and $\beta =
2\sqrt{2}G_F/\hbar$ for neutrinos of the same flavor.   

Suppose now that the neutrino pulses are coherent and uni-directional.
If one species, the pump species (denoted by the index $p$), 
dominates the number density as compared to the signal species
(denoted by the index $s$), i.e. $n_p \gg n_s$, we obtain the equation
\numparts
\begin{equation}
  i\frac{\partial{\mnp}_s}{\partial t} +
  \frac{\alpha}{2}\frac{\partial^2{\mnp}_s}{\partial z^2} +
  \beta|{\mnp}_p|^2{\mnp}_s = 0 ,
\label{split}
\end{equation}
for the signal wave function ${\mnp}_s$, where the driving nonlinear
term is determined from 
\begin{equation}
  i\frac{\partial{\mnp}_p}{\partial t} +
  \frac{\alpha}{2}\frac{\partial^2{\mnp}_p}{\partial z^2} -
  \beta|{\mnp}_p|^2{\mnp}_p = 0 .
\label{dark}
\end{equation}
\label{system}
\endnumparts
It is well known that Eq.~(\ref{dark}) have modulationally stable dark
soliton solutions \cite{Kivshar}, of which the fundamental soliton is 
\begin{equation}
  {\mnp}_p = {\mnp}_0\tanh\left(
  \sqrt{\beta/\alpha}\,{\mnp}_0z 
  \right)\,\exp({-i\beta{\mnp}_0^2t}) .
\label{fundamental}
\end{equation} 

\begin{figure}
\begin{center}
  \includegraphics[width=.6\textwidth]{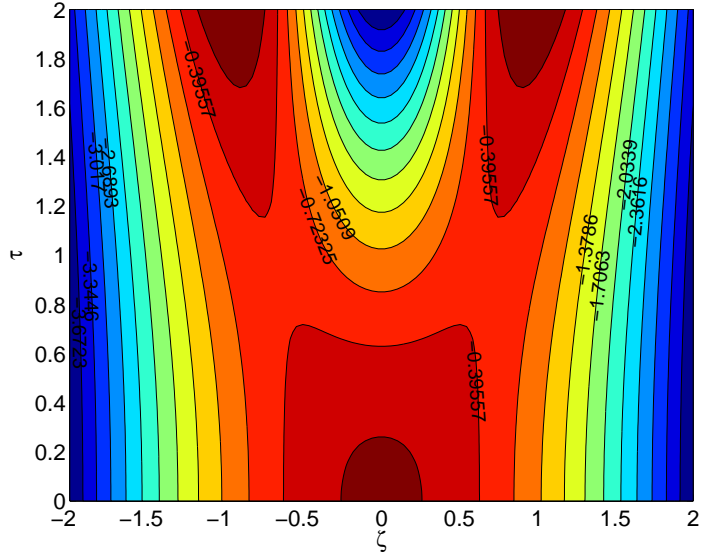}
\end{center}
  \caption{Contour plot of $\ln[n_s(t,z)/n_{s0}]$ ($b = 1$) for the
  splitting case.}  
  \label{split_contour}
\end{figure}

\begin{figure}
\begin{center}
  \includegraphics[width=.6\textwidth]{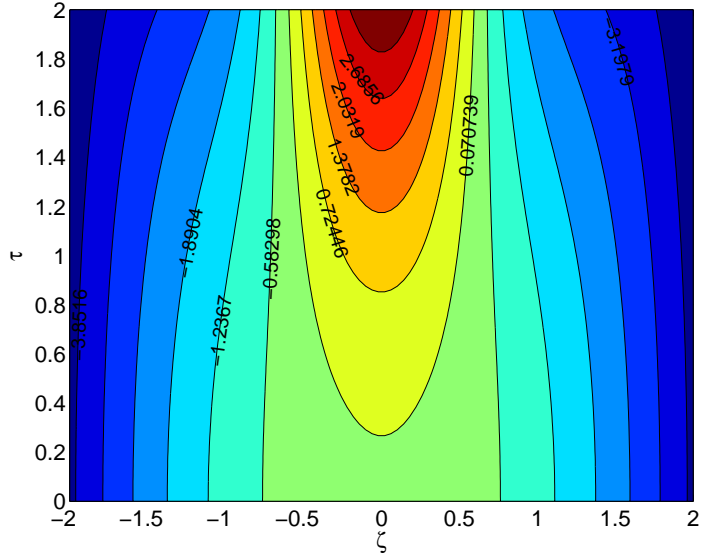}
\end{center}
  \caption{Contour plot of $\ln[n_s(t,z)/n_{s0}]$ ($b = 1$) for the
  focusing case.}  
  \label{focus_contour}
\end{figure}

Following Ref.~\cite{Helczynski-etal}, we perform a perturbation
analysis of Eq.~(\ref{split}) for a given solution ${\mnp}_p(t,z)$ to 
Eq.~(\ref{dark}). Separating ${\mnp}_s(t,z)$ into its real amplitude
$\sqrt{n_s(t,x)}$ and phase $\theta_s(t,z)$, we have, to lowest order
in $t$, 
\begin{equation}
  \theta_s(t,z) \approx \beta\int_0^t |{\mnp}_p(t',z)|^2\,dt' \approx
  \beta |{\mnp}_p(0,z)|^2 t,
\end{equation}
where the last equality follows if the interaction takes place over a
time-scale short compared to the time over which the pump amplitude
changes. The signal number density $n_s(t,z)$ is then  
\begin{equation}
\fl  n_s(t,z) \approx n_s(0,z)\exp\bigg[ -\frac{\alpha\beta
  t^2}{2}\bigg( 
  \frac{d^2 n_p(0,z)}{dz^2}
  + \frac{d n_p(0,z)}{dz}\frac{d}{dz}\ln
  n_s(0,z) \bigg) \bigg] ,
\end{equation}
where $n_p(t,z) = |{\mnp}_p(t,z)|^2$ is the pump number density. Using
the fundamental soliton (\ref{fundamental}) as the pump pulse, with
${\mnp}_0^2 = n_p(0,0) \equiv n_{p0}$, and a Gaussian distribution for
the initial signal pulse,
\begin{equation}
  n_s(0,z) = n_{s0}\exp\left(-z^2/a^2\right),
\label{gauss}
\end{equation}
where $a$ is the width and $n_{s0} = n_s(0,0)$, we find that 
\begin{equation}
\fl  n_s(t,z) = n_{s0}\exp\Big\{ -b^2\zeta^2
  -\Big[ \left(3\sech^2\zeta - 2
  \right)\!\sech^2\zeta 
  - 
  2b^2\tanh\zeta\sech^2\zeta \Big]\tau^2  \Big\} ,
\label{solution}  
\end{equation}
where $\zeta = z/z_0$, $\tau = \alpha t/z_0^2$, $b = z_0/a$, and $z_0 =
(\beta n_{p0}/\alpha)^{1/2}$. We note that 
$b$, $\tau$ and $\zeta$ are dimensionless. The solution
(\ref{solution}) is plotted in
Fig.~\ref{split_contour}, where the splitting of the collective
neutrino state can clearly be seen.

In the case of neutrino--neutrino or anti-neutrino--anti-neutrino 
interaction, Eq.~(\ref{interaction}) will be modified according to 
\begin{eqnarray}
  i\frac{\partial{\mnp}_{i}}{\partial t} +
  \frac{\alpha}{2}\nabla^2{\mnp}_{i} -
  \beta(\langle|{\mnp}_{1}|^2\rangle + \langle|{\mnp}_{2}|^2\rangle
  ){\mnp}_{i} = 0 ,   
\label{interaction2}
\end{eqnarray}
where $i= 1, 2$. For a dark soliton pump beam (\ref{fundamental}) and 
a Gaussian signal beam (\ref{gauss}), the approximate dynamics,
analogous to the neutrino--anti-neutrino interaction, can be obtained
from (\ref{solution}) by letting $\tau^2 \rightarrow -\tau^2$. The 
result is seen in Fig.~\ref{focus_contour}, where the focusing is
depicted.  

The above discussion is based on the approximate solution
(\ref{solution}). However, numerical studies of the system (\ref{split})
and (\ref{dark}) show the same behaviour, when (\ref{fundamental})
and (\ref{gauss}) are used as initial conditions. Thus, a more
complete solution supports the notion that the
approximate analytical expression (\ref{solution}) gives an accurate
description of the dynamics of interacting neutrinos and
anti-neutrinos.  

In conclusion, we have considered the nonlinear interactions between 
neutrinos and anti-neutrinos, mediated by a dense
neutrino background of the same flavour.
It is shown that the self-interaction between neutrinos and
anti-neutrinos comes about due to their coupling via the weak 
nuclear force. Such interactions are governed by coupled cubic
Schr\"odinger 
equations for the slowly varying envelopes of the neutrino and
anti-neutrino  
wave functions. Analytical and numerical analysis of the 
multi-dimensional cubic Schr\"odinger equations reveal that a
collective  
neutrino state can be split or focused 
due to the presence of a dense soliton background. The present results  
suggest 
that nonlinear excitations in neutrino plasmas can significantly
affect the propagation of intense neutrino and anti-neutrino beams.
The neutrino beam splitting can account for the energy loss of 
neutrinos, while the self-focusing renders neutrino burst 
propagation over a long distance.

\end{document}